\documentclass[12pt]{article}
\usepackage{amsmath,amsfonts,amssymb}
\usepackage[T1]{fontenc}
\usepackage{authblk}

\newcommand{\cH}{{\mathcal H}}
\newcommand{\cB}{{\mathcal B}}
\newcommand{\ZZ}{{\mathbb Z}}
\newcommand{\RR}{{\mathbb R}}
\newcommand{\CC}{{\mathbb C}}
\newcommand{\End}{{\rm End}}
\newcommand{\ch}{{}{\rm ch}_{>0}}
\newcommand{\ra}{\rightarrow}
\renewcommand{\Im}{{\rm Im}}
\newcommand{\hH}{{\hat H}}

\begin{document}

\title{Localization properties of Chern insulators}

\author{Roman Bezrukavnikov}
\affil{Massachusetts Institute of Technology}
\author{Anton Kapustin}
\affil{California Institute of Technology}

\maketitle

\abstract{We study the localization properties of the equal-time electron Green's function in a Chern insulator in an arbitrary dimension and with an arbitrary number of bands. We prove that the Green's function cannot decay super-exponentially if the Hamiltonian is finite-range and the quantum Hall response is nonzero. For a general band Hamiltonian (possibly infinite-range), we prove that the Green's function cannot be finite-range if the quantum Hall response is nonzero. The proofs use methods of algebraic geometry.}
\section{Introduction}

Quantum Hall Effect remains one of the most spectacular manifestations of the importance of topology in quantum physics. Integer QHE is observed in 2D materials in high magnetic fields, but it has been realized early on that the role of the magnetic field is mainly to induce a strong breaking of time-reversal symmetry, and that similar physics can in principle occur in any insulating material when this symmetry is broken \cite{Haldane}. Such hypothetical materials are dubbed Chern insulators, since the relevant topological invariant is the 1st Chern class of the vector bundle formed by the wavefunctions of electrons in the valence band. The base of this bundle is a real torus (the space of quasi-momenta).

While topology has found many applications in condensed matter physics, algebraic geometry has not been used much so far. In this note we explain how a well-known result about algebraic vector bundles over algebraic tori can be used to prove  interesting no-go theorems about Chern insulators. 

Let $B$ be a finite-dimensional Hilbert space (the band space). The Hilbert space of a single electron in a $d$-dimensional crystal with band space $B$ is $\cH(B)=L^2(T^d,B)$, where $T^d=\RR^d/(2\pi \ZZ)^d$ is the $d$-dimensional torus (the Brillouin zone). By Fourier transform, we can also identify $\cH(B)$ with $\ell^2(\Lambda,B)$, where $\Lambda=\ZZ^d\subset \RR^d$ is the lattice of characters for $T^d$. 

A Hamiltonian for a single-electron system with band space $B$ is a continuous function $H:T^d\ra \End(\cB)$ satisfying the Hermiticity condition $H(p)^\dagger=H(p)$. Here $p$ is an affine coordinate on $T^d$. Fourier transform maps $H(p)$ to an operator $\hH$ on $\ell^2(\Lambda,B)$ with matrix elements
\begin{equation}
\langle \lambda\vert\hH\vert\lambda'\rangle=\int \frac{d^dp}{(2\pi)^d} H(p) e^{i\langle\lambda-\lambda',p\rangle}
\end{equation}

A Hamiltonian $H(p)$ is said to be finite-range if it has the form
\begin{equation}
H(p)=\sum_{\lambda\in\Lambda_0} h(\lambda) e^{-i\langle\lambda,p\rangle},
\end{equation}
where $\Lambda_0$ is a finite subset of $\Lambda$ symmetric with respect to the origin, and the matrices $h(\lambda)\in \End(B)$ satisfy
$h(\lambda)^\dagger=h(-\lambda)$. The last condition ensures that $H(p)^\dagger=H(p)$. The name ``finite-range'' refers to the fact that the corresponding operator $\hH$ on $\ell^2(\Lambda,B)$ has matrix elements
\begin{equation}
\langle \lambda\vert\hH\vert\lambda'\rangle=\sum_{\lambda-\lambda'\in \Lambda_0} h(\lambda-\lambda'),
\end{equation}
and thus for a fixed $\lambda$ is nonzero only for a finite set of values of $\lambda'$.

Since the function $H(p)$ is Hermitian, its eigenvalues are real for all $p$. A Hamiltonian $H(p)$ is called gapped if its eigenvalues are nonzero everywhere on $T^d$. In this note we will only consider gapped Hamiltonians. Let $P_+(p)$ and $P_-(p)$ denote projectors to the positive and negative eigenspaces of $H(p)$, respectively. Their images define a pair of complex vector bundles over $T^d$ which we denote $\cB_+$ and $\cB_-$. If we denote by $\cB$ the trivial vector bundle over $T^d$ with fiber $B$, then clearly $\cB_+\oplus \cB_-=\cB$. Therefore the Chern characters of $\cB_+$ and $\cB_-$ sum up to $\dim B$. The Chern character of $\cB_-$ controls the quantum Hall response of the system and thus is an important physical quantity \cite{TKNN}. More precisely, it is the component of the Chern character which sits in degree higher than zero that determines the quantum Hall response. We will denote it $\ch(\cB_-)=-\ch(\cB_+)$. If $\ch(\cB_\pm)$ is nonzero, one says that the gapped Hamiltonian $H(p)$ describes a Chern insulator. The most physically interesting case is $d=2$, in which case $\ch(\cB_\pm)$ reduces to the 1st Chern class $c_1(\cB_\pm)$. For a review of Chern insulators and other topological insulators see \cite{BernevigHughes}.

The spectral projector $P_-(p)$ also controls the localization properties of the electrons in the state obtained by filling all negative-energy bands. Let us denote by $G(\lambda-\lambda',t-t')$  the 2-point Green's function. Then its value at $t=t'$ is given by the Fourier transform of $P_-(p)$:
\begin{equation}
G(\lambda-\lambda',0)=\int \frac{d^d p}{(2\pi)^d} 
e^{i\langle\lambda-\lambda',p\rangle} P_-(p).
\end{equation}
The decay rate of $G(\lambda-\lambda',0)$ for large $|\lambda-\lambda'|$ is determined by the analytic properties of $P_-(p)$ and vice versa. For example, $P_-(p)$ is a $C^\infty$ function of $p$ if and only if $G$ decays faster than any power of $|\lambda-\lambda'|$. In fact, from the formula
\begin{equation}
P_-(p)=\frac{1}{2\pi i}\oint_{C_-} \frac{dE}{E-H(p)},
\end{equation}
where $C_-$ is a closed contour in the complex $E$-plane  surrounding the set of negative eigenvalues of $H(p)$, we see that $P_-(p)$ is real-analytic if $H(p)$ is gapped and real-analytic.\footnote{Real-analyticity of $H(p)$ is equivalent to the requirement that the matrix elements  $\langle \lambda\vert \hH\vert\lambda'\rangle$, $\lambda,\lambda'\in\Lambda$, decay exponentially as $|\lambda-\lambda'|\ra\infty$.} Since $T^d$ is compact, a real-analytic projector $P_-(p)$ can be analytically continued to a finite-width neighborhood of $T^d$ in its complexification $(\CC^*)^d$. Here $\CC^*=\CC\backslash\{0\}$. Therefore the equal-time Green's function $G$ corresponding to a real-analytic gapped Hamiltonian decays at least exponentially:
\begin{equation}\label{expdecay}
||G(\lambda-\lambda',0)||\leq g e^{-|\lambda-\lambda'|/\ell}
\end{equation}
for some constants $g, \ell>0$. In such a case one says that the correlation length is at most $\ell$. The lower bound of the set of all $\ell$ for which (\ref{expdecay}) holds is a non-negative number called the correlation length of the free system described by $H(p)$. For any real-analytic gapped $H(p)$, the correlation length is finite, as we just showed.

The question we want to address is if non-zero $\ch(\cB_\pm)$ is compatible with the equal-time Green's function decaying faster than any exponential. In other words, can a Chern insulator have zero correlation length? In particular, can the Green's function $G(\lambda-\lambda',0)$ of a Chern insulator be finite-range, i.e. vanish for $|\lambda-\lambda'|>R$ for some $R$? Of course, there are trivial examples of gapped Hamiltonians where this is true: just take $H(p)$ to be a constant non-degenerate Hermitian matrix $h_0$, so that $P_-(p)$ is a constant matrix as well, and $G(\lambda-\lambda',0)$ is non-zero only for $\lambda=\lambda'$. But in this case both $\cB_+$ and $\cB_-$ are topologically trivial, so $\ch(\cB_\pm)=0$.

There is a heuristic reason why a non-zero $\ch(\cB_\pm)$ might be incompatible with zero correlation length. It is well-known that a Chern insulator has gapless excitations along a spatial boundary \cite{BernevigHughes}. Thus the correlation length is infinite at the boundary. A scenario where the correlation length is zero in the bulk and infinite at the boundary seems implausible.

For a general gapped Hamiltonian, we  show that a finite-range equal-time Green's function implies $\ch(\cB_\pm)=0$. Thus completely localized electrons are incompatible with a nonzero quantum Hall response.

For a finite-range gapped Hamiltonian, we prove a stronger result: zero correlation length implies vanishing $\ch(\cB_\pm)$. Thus super-exponential localization is incompatible with a nonzero quantum Hall response, provided the Hamiltonian is finite-range.

Despite the above no-go results, it might be possible to construct an infinite-range gapped Hamiltonian with a zero correlation length and a non-vanishing quantum Hall response. Such a system would have the peculiar property that the correlation length is zero in the bulk and infinite at the boundary.

The authors would like to thank Alexei Kitaev for bringing the problem discussed in this paper to their attention.
The research of A.\ K.\ was supported in part by the U.S.\ Department of Energy, Office of Science, Office of High Energy Physics, under Award Number DE-SC0011632. A.\ K.\ was also supported by the Simons Investigator Award. The work was completed at the Aspen Center for Physics which is supported by the National Science Foundation grant PHY-1607611. R.\ B.\ was partly supported by the NSF grant DMS-1601953. 

\section{General Hamiltonian}

Suppose there exist $g>0$ and $a>0$ such that
\begin{equation}\label{expdecay}
||G(\lambda,0)||<g \exp(-a|\lambda|).
\end{equation}
Then the Fourier series 
$$
P_-(p)=\sum_{\lambda\in\ZZ^d} e^{-i\langle\lambda,p\rangle} G(\lambda,0)
$$
is an analytic function of $p_1,\ldots,p_d$ in a strip $|\Im\, p_i|<a$, $i=1,\ldots,d$. If the estimate (\ref{expdecay}) holds for all $a>0$, then $P_-(p)$ is an entire function of $p_1,\ldots,p_d$. Since it is also periodic in $p_1,\ldots,p_d$ with period $2\pi$, it can be regarded as an analytic function on $(\CC^*)^d$ with coordinates $w_k=\exp(i p_k)$, $k=1,\ldots,d$, $w_i\neq 0$. Here $\CC^*=\CC\backslash\{0\}$. 

Suppose furthermore that $G$ is finite-range, i.e. there exists $R>0$ such that $|\lambda|>R$ implies $G(\lambda,0)=0$. Then matrix elements of $P_-(p)$ are bounded from above by a multiple of 
$$
R^d\exp\left(R\sum_{i=1}^d |{\rm Im}\, p_i|\right)=R^d \prod_{i=1}^d |w_i|^{\pm R}.
$$
(This is the ``easy'' direction of Paley-Wiener-type theorems, see e.g. \cite{SteinWeiss}). Since $P_-$ is also an analytic function of $w_1,\ldots,w_d$, Liouville's theorem from complex analysis implies that $P_-$ is a Laurent polynomial in $w_1,\ldots, w_d$.

The bundle $\cB_-$ is the image of $P_-$ and thus extends to an analytic vector bundle on $(\CC^*)^d$. Since $P_-$ is an algebraic matrix-valued function, $\cB_-$ is an algebraic vector bundle. 

While any topological vector bundle on $(\CC^*)^d$ admits a (unique)  analytic structure\footnote{This follows from a theorem of H. Grauert \cite{GrauertI, GrauertII} (see also Theorem 5.3.1 in \cite{Steinmanifolds}) and the fact that $(\CC^*)^d$ is a Stein manifold.}, any algebraic vector bundle on $(\CC^*)^d$ is trivial. This is a special case of a theorem of I. Gubeladze \cite{Gubeladze} (see also Corollary V.4.10 in \cite{Lam}) generalizing the famous result of D. Quillen \cite{Quillen} and A. Suslin \cite{Suslin} that all algebraic vector bundles on $\CC^d$ are trivial. Thus $\cB_-$ is topologically trivial, and $\ch(\cB_\pm)=0$.

Triviality of algebraic vector bundles over $\CC^d$ and $(\CC^*)^d$ is a deep fact. 
For the most physically interesting case $d=2$, it is sufficient to show that $c_1(\cB_\pm)=0$. Since $c_1$ of a vector bundle equals $c_1$
of its determinant line bundle, this fact follows from triviality
of algebraic line bundles on $X$, which is a direct consequence
of the unique factorization property of the ring of Laurent polynomials in several variables (recall that isomorphism classes of algebraic line 
bundles are in bijection with the divisor classes). Alternatively, one
can use the fact that the closure of a divisor on $(\CC^*)^d$ regarded as a sub-variety of $\CC^d$ is a divisor on $\CC^d$, thus an algebraic line bundle on $X$ is a restriction of a line bundle on $\CC^d$, hence has trivial Chern class (here we used smoothness of $\CC^d$ to identify 
Cartier (locally principal) and Weil divisors on it, see e.g. \cite[Chapter 3.1]{Shafarevich}). 
(For a more detailed argument along these lines, see the proof of Theorem II.1.3 in \cite{Lam}). 

\section{Finite-range Hamiltonian}

Now suppose that $H(p)$ is finite-range, while
$G(\lambda-\lambda',0)$ decays faster than any exponential. We have seen in the previous section that the latter property implies that $P_-$ is an analytic function on $(\CC^*)^d$. We will now show that under this weaker condition
on $P_-$ we still have $\ch(\cB_\pm)=0$ provided that the Hamiltonian is finite range.  

{\bf Lemma} {\em Let $H$ be an algebraic function on $X=(\CC^*)^d$ with values in $Mat_n(\CC)$. As before, we assume
that $H(s)$ is a nondegenerate Hermitian matrix when $s\in T^d$, so the bundles $\cB_\pm$ on the compact torus $T^d$ and projectors $P_\pm:T^d\to Mat_n(\CC)$ are well defined.

a) If $P_\pm$ admits an analytic extension to $X$ then the bundles $\cB_\pm$ extend to an algebraic vector bundle on $X$,
hence $\cB_\pm$ is a trivial vector bundle.

b) If $P_\pm$ admits a meromorphic extension to $X$ then the determinant line bundles $det(\cB_\pm)$ 
extend to an algebraic line bundle on $X$,
hence $c_1(\cB_\pm)=0$.}

{\em Proof.}
a)  The analytic continuation of the Hamiltonian $H(p)$ to $(\CC^*)^d$ is a regular algebraic function with values in $\End(B)$. By definition, $H(p)$ commutes with $P_\pm(p)$ and thus their analytic continuations also commute. Hence $H$, regarded as an endomorphism of the trivial vector bundle $\cB=\cB_+\oplus\cB_-$, preserves both sub-bundles. Let $H_\pm$ denote the restrictions of $H$ to $\cB_\pm$. These are analytic endomorphisms of analytic bundles $\cB_\pm$. The characteristic polynomial 
\begin{equation}
\Delta(t)=\det(t-H)
\end{equation}
decomposes as a product of characteristic polynomials of $H_\pm$:
\begin{equation}
\Delta(t)=\Delta_+(t)\Delta_-(t).
\end{equation}
$\Delta_\pm(t)$ are monic polynomials in $t$ whose coefficients are analytic functions on $(\CC^*)^d$. On the other hand, since $H$ is algebraic, $\Delta(t)$ is a monic  polynomial in $t$ whose coefficients are regular algebraic functions on $(\CC^*)^d$. Liouville's theorem then implies that both $\Delta_+(t)$ and $\Delta_-(t)$ have regular algebraic coefficients.

Consider $\Delta_+(H)$. It is an algebraic endomorphism of $\cB$ which preserves the decomposition $\cB=\cB_+\oplus \cB_-$. It vanishes on $\cB_+$ (thanks to the Cayley-Hamilton theorem). Also, it is non-degenerate on $\cB_-$ away from points where a root of $\Delta_+(t)$ is equal to a root of $\Delta_-(t)$. Such ``bad'' points form an algebraic divisor $R$ in $(\CC^*)^d$ (given by the vanishing of the resultant of the polynomials $\Delta_+(t)$ and $\Delta_-(t)$). Thus when we restrict $\Delta_+(H)$ to $(\CC^*)^d\backslash R$, it becomes an endomorphism of $\cB$ which annihilates $\cB_+$ and is non-degenerate on $\cB_-$. Thus the restriction of $\cB_-$ to $(\CC^*)^d\backslash R$ can be identified with the image of $\Delta_+(H)$ and therefore is an algebraic sub-bundle of the trivial vector bundle $\cB$ on $(\CC^*)^d\backslash R$.

We showed that $\cB_-$ is an analytic vector bundle on $(\CC^*)^d$ which is  algebraic on the Zariski-open subset $(\CC^*)^d\backslash R$. A subbundle of a trivial bundle corresponds to a map from the base to the Grassmann variety.
By basic algebraic geometry, an analytic map of complex algebraic manifolds which is algebraic on a Zariski open dense subset is algebraic; thus $\cB_\pm$  is algebraic on the whole $(\CC^*)^d$. 

To check (b) observe that the factorization $\Delta(t)=\Delta_+(t)\Delta_-(t)$ exists by the above argument
even if the projectors $P_\pm$ are only assumed to extend to $X$ meromorphically. Thus the vector bundles $\cB_\pm$
extend to algebraic vector bundles on the complement to $R$. Since a rational map from a normal variety to a projective
one (in particular, to a Grassmann variety) extends to a complement of a subvariety of codimension at least two, we 
see that $\cB_\pm$ extend to algebraic bundles on $X\setminus S$, ${\mathrm {co}}\dim(S)\geq 2$. As was pointed out above,
for such an $S$ the Picard group $Pic(X\setminus S)=Pic(X)$ is trivial, thus the line bundles $det(\cB_\pm)$ are trivial.
$\square$

Part (b) of the Lemma shows that $c_1(\cB_\pm)=0$ for a finite range Hamiltonian even if $G(\lambda-\lambda',0)$
satisfies a weaker assumption than the faster than exponential decay: the latter is equivalent to existence of an analytic continuation of $P_\pm$, while here we only need to know existence of a meromorphic continuation. However, the physical
or analytic significance of that weaker condition remains to be understood.

\end{document}